\documentclass[prl,floatfix,superscriptaddress,twocolumn,aps,showpacs,preprintnumbers,amsmath,amssymb]{revtex4}
\usepackage[latin1]{inputenc}

\usepackage{graphicx}% Include figure files
\usepackage{psfrag}
\usepackage{wasysym}
\newcommand{\ColorOnline}{(Color online) }

\newcommand{\ket}[1]{\ensuremath{\vert{#1}\rangle}}
\newcommand{\bra}[1]{\ensuremath{\langle{#1}\vert}}
\newcommand{\e}{\ensuremath{\mathrm{e}}}

\newcommand{\be}{\begin{equation}}
\newcommand{\ee}{\end{equation}}
\newcommand{\bea}{\begin{eqnarray}}
\newcommand{\eea}{\end{eqnarray}}

\newcommand{\HH}{{\ensuremath{\cal H}}}

\def\bea{\begin{eqnarray}}
\def\eea{\end{eqnarray}}
\def\ben{\begin{equation}}
\def\een{\end{equation}}
\def\benu{\begin{enumerate}}
\def\enu{\end{enumerate}}

\def\n{n}

\def\sss{\scriptscriptstyle\rm}

\def\1var{(\bx_1...\bx\N)}

\def\br{{\bf r}}

\def\bx{{\br t}}

\def\bj{{\bf j}}

\def\bE{{\bf E}}

\def\s{_{\sss S}}
\def\xc{_{\sss XC}}

\def\N{_{\sss N}}

\def\ee{_{\rm ee}}

\def\sph_int{ {\int d^3 r}}

\input rotate

\def\tot{_{\text{tot}}}

\begin{document}

%\preprint{Schmitteckert \& Evers, DFT$\oplus$DMRG/TKM}

\title{Exact ground state density functional theory for impurity
  models coupled to external reservoirs and transport calculations}

\author{Peter Schmitteckert}%
  \affiliation{\mbox{Institut f\"ur Theorie der Kondensierten Materie, Universit\"at Karlsruhe, 
    D-76128 Karlsruhe, Germany}} %% box for having no linebreak
\author{Ferdinand Evers}%
  \affiliation{\mbox{Institut f\"ur Theorie der Kondensierten Materie, Universit\"at Karlsruhe, 
    D-76128 Karlsruhe, Germany}} %% box for having no linebreak
  \affiliation{Institut f\"ur Nanotechnologie, 
    Forschungszentrum Karlsruhe, D-76021 Karlsruhe,
    Germany}

\date{\today}

\begin{abstract}
A method is presented, which employs the density matrix renormalization group
technique in order to construct exact ground state 
exchange correlation functionals for models of 
correlated electron systems coupled to external
reservoirs. The technique is applied to the $M$-site resonant level
model. We calculate its exact Kubo-conductance, 
which is  available within DMRG, and compare to the single-particle 
conductance obtained from Kohn-Sham energies and orbitals of the exact
ground state density functional theory (DFT). 
It is found that the position of transport resonances is reproduced
essentially exactly, while deviations in the level broadening can be
less than 1\%  and do not exceed 10\%.
Our findings lend strong support to a recently held point 
of view, namely that approximations in the ground state functionals used
in DFT based transport calculations
can lead to drastic errors in transport calculations while exchange
correlation contributions to the induced effective potential 
tend to be less significant.

\end{abstract}

%%\pacs{72.15.Rn, 05.45.Df}
\pacs{31.15.Ar, 71.15.Mb, 81.07.Nb}

\maketitle

%\section{Introduction}

%\paragraph{Introduction:}

Running an electrical current through individual molecules and 
being able to control the current flow by molecular design 
is one of the intriguing aspects of Molecular Electronics. 
Naturally, the interesting behavior of such ``devices'' is 
frequently related to specific properties of the molecule at hand. 
Therefore, in order to properly understand experimental findings,
notably the current voltage characteristics, ab initio
calculations play a key role in this field. 

The challenge in such ab initio calculations for transport properties
is to accurately describe interaction effects on the
junction together with wavefunction hybridization with the contacts. 
To deal with the latter, it is mandatory to treat a sizable part of
the contact on the same footing as the molecule. 
To achieve sufficient system sizes, the current``standard method'' is based on the 
density functional theory (DFT).\cite{brandbyge02,evers03prb,arnold07}

A justification of the principal approach, namely the use of the single particle
Keldysh formalism in conjunction with the Kohn-Sham (KS) scheme of  
time dependent (TD) DFT, has been proposed in Ref. 
\cite{evers03prb} on a heuristic level and by Stefanucci and Almbladth
has been based on a detailed analysis using the Keldysh-technique.\cite{stefanucci:195318}.

In practical calculations approximate 
exchange correlation (XC) functionals have to be employed. 
Consequences of such approximations 
have been investigated in Ref. \cite{koentopp:121403}, 
where a relation between the linear 
current $\bj$, the KS-Kubo-conductivity and the TDDFT 
functional $\bE\xc(\br\omega)$ has been derived:
\ben
\bj(\br\omega) = \int d^3r'\; \hat \sigma\s[\n_0] (\br\br'\omega)\;
\left(\bE\tot(\br'\omega) + \bE\xc(\br'\omega)\right).
\label{e1}
\een
($\bE\tot(\br\omega)$ denotes the sum of external and Hartree field, 
$\n_0({\bf r})$ the ground state electron density.) 
%{\bf FE: nKS}
In one dimension and the {\it dc}-limit, $j({\bf r},\omega\to 0){=}I$, 
so Eq. (\ref{e1}) reduces to 
%\ben
$ I = g\s \left( V_\text{tot} + V_\text{XC} \right)$
%\label{e1b}
%\een
introducing the KS-conductance, $g\s$, the physical voltage $V_\text{tot}$ and 
its XC-shift $V_\text{XC}$ \cite{koentopp:121403}.

In most current implementations of the standard method, 
the linear current response derives from calculating $g\s$ 
in a local density approximation. Moreover, 
%(analogous to Eq. (\ref{e8}));
the XC-terms, $V\xc$ or $\bE\xc$, are universally ignored even though
$\bE\xc(\omega)$ is generally important for the calculation of
excitation energies \cite{dreizler90}.
Such approximations must be expected to have severe consequences. 
Most notably, the missing derivative
discontinuity in available XC-functionals 
can artificially boost the DFT level broadening $\Gamma_{\rm DFT}$
observed in the variation of $g\s$ with the gate voltage, $V_\text{gate}$,
from its true value $\Gamma$ up to the interaction energy 
$U\gg \Gamma$ and thus completely impair 
an accurate prediction of transport coefficients \cite{thesisAA07}. 
This effect is a major suspect to cause much of the strong
deviations that are frequently observed between theoretical and experimentally
determined transport coefficients.\cite{koentopp:121403,TFSB05,ke06}
%%
%% our stuff 
%%

Further analysis of Eq. (\ref{e1}) has been impaired by the fact, that 
for correlated molecular systems, which have been coupled to external
reservoirs, exact XC functionals are not available. 
To overcome this fundamental barrier, we put forward a general idea in this work, 
namely to marry DFT with the 
density matrix renormalization group (DMRG) technique. Our approach
allows to construct {exact} ground state (GS)
functionals for a broad class of generic model systems
of correlated fermions {including external reservoirs}. 
Here, we apply this method to obtain the exact KS conductance $g\s$. 
By comparing to the  exact physical conductance, $g=I/V_\text{tot}$, 
it is demonstrated that $g\s$ gives a highly accurate estimate for position and 
broadening of transport resonances implying that dynamical corrections
$V_\text{XC}$, including the viscous part \cite{sai05},
are small. {We show this specifically for the
interacting resonant level model (IRLM) and the resonant chain model
under conditions of the Coulomb blockade. 
Arguments are given, why we expect our results to be
more generally valid.}

\begin{figure}[t]
\begin{center}
    \psfrag{R}{{$\cal R$}}
    \includegraphics[width=0.475\textwidth]{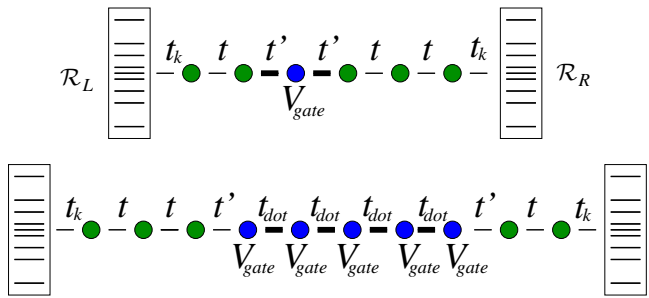} %{MomSetup.v3.eps}
    \caption{\ColorOnline Schematic representation of the
    calculational setup. Filled circles indicate the extended
    molecule, $e\!\mathcal{M}$.
    Gate voltage, $V_\text{gate}$, is applied to molecule (dot),
    $\mathcal{M}$, only. Interacting sites are coupled
    by fat line links. Link labels indicate hopping amplitudes
    in the leads $t$, on the molecule $t_\text{dot}$ and near the
    contacts $t'$ and $t_k$. (${t{=}1,t_\text{dot}{=}0.5}$ in all
    calculations presented.) $\mathcal{R}_L,\mathcal{R}_R$ denotes the
    reservoirs, here in $k$-space representation. 
    Upper panel: single site resonance level model. Lower
    panel: five site model.\label{Fig:SiteLayout}}
\end{center}
\end{figure}

%\paragraph{Model definition:}
The IRLM and its extension including reservoirs, that we employ in this
work, is a suitable testbed to study strong correlation effects on 
transport. The model Hamiltonian reads 
$\HH = \HH_{e\mathcal{M}}{+}\HH_U{+}\HH_{\mathcal{R}}{+}\HH_{T}$, 
with 
\begin{eqnarray}
\label{e2}
    \HH_{e\mathcal{M}} &=& V_\text{gate} \sum_{\ell\in \mathcal{M}} \hat{c}_\ell^\dagger  \hat{c}^{}_\ell
       \,-\, \!\!\!\!\sum_{\ell,\ell-1\in e\!\mathcal{M}} \!\!\!\!
       \big(t_\ell \hat{c}_\ell^\dagger \hat{c}^{}_{\ell-1}+ \text{h.c.}\big)\\
    \HH_U &=& U \sum_{\ell,\ell-1\in \mathcal{M}} 
      \Big( \hat{n}_\ell-\frac 12\Big) \Big( \hat{n}_{\ell-1} - \frac 12\Big),
\end{eqnarray}
where $\hat{c}^\dag_\ell$ ($\hat{c}^\dag_k$) and $\hat{c}^{}_\ell$ ($\hat{c}^{}_k$)  are spinless
fermionic creation and annihilation operators at site $\ell$ (at momentum $k$),
$\hat{n}_\ell = \hat{c}^\dag_\ell \hat{c}^{}_\ell$. 
Furthermore, $\HH_{\mathcal{R}}{=} \sum_{k\in \mathcal{R}_L,\mathcal{R}_R}
\epsilon_k \hat{c}_k^\dagger  \hat{c}^{}_k$ and  
\begin{eqnarray}
%    \HH_{\mathcal{R}} &=& \sum_{k\in \mathcal{R}_L,\mathcal{R}_R} \epsilon_k \hat{c}_k^\dagger  \hat{c}^{}_k,\\
    \HH_{T} &=& -\Big( \sum_{k\in \mathcal{R}_L} t_k \hat{c}_{k}^\dagger  \hat{c}^{}_{1}
            \,+\,   \sum_{k\in \mathcal{R}_R} t_k \hat{c}_{k}^\dag  \hat{c}^{}_{M_E} \Big) \,+\, \text{h.c.}.
\end{eqnarray}
Indices denote Hilbert spaces 
of the molecule proper (nanostructure/quantum dot), $\mathcal{M}$,
of the extended molecule, $e\!\mathcal{M}$ and of the
left and right electrode reservoirs, $\mathcal{R}_L, \mathcal{R}_R$. $\HH_T$ denotes 
the tunneling Hamiltonian describing the contact between reservoirs and the
extended molecule; the interaction on the molecule is modeled by
$\HH_U$. ($U{=}2t$ in all calculations presented.)
The indices $1$ and $M_E$ denote the first and last site in $e\!\mathcal{M}$. The
general setup is displayed in Fig.~\ref{Fig:SiteLayout}.
For the single level model the interaction $U$ extends on the link between the
resonant level and the neighboring lead sites.
In all calculations we have considered the case of a half filled
band and zero temperature, $E_{\rm Fermi}{=}0, T=0$. 

%\paragraph{Exact ground state with DMRG:}
We calculate the GS  of our model Eq.~(\ref{e2}) 
by means of a DMRG calculation. DMRG\cite{White:1992,DMRG:Reviews}
is a method that searches for an optimized subspace of the complete Hilbert space
in which selected many body states can be described accurately.
Notice, that our setup include the leads in 
momentum space representation, $\HH_{\mathcal{R}}$, which is non-standard, but
crucial for later transport calculations, Fig.~\ref{Fig:SiteLayout}; for details 
see.\cite{Bohr_Schmitteckert:2007}
In this way we obtain the local electron 
density on $e\mathcal{M}$ together with the occupation number of lead levels. 
Typically, we use more than 1300 states per block and ten finite lattice sweeps. 

%\paragraph{Exact DFT:}
Next, we briefly explain how an exact DFT may be constructed 
generalizing earlier ideas 
by Gunnarson and Sch\"onhammer.\cite{Gunnarson_Schoenhammer:1986}
We define a 
KS-Hamiltonian 
%\ben
$\HH\s = \HH_0 + V_{\text{HXC}},$
%\een
with 
$\HH_0 {=} \HH_{e\mathcal{M}}{+}  \HH_{\mathcal{R}} {+} H_{T}$
(free fermions) and an XC potential
$V_{\text{HXC}} {=} \sum_j v_j \hat{n}_j \,$
%\begin{equation}
%\label{e6}
%	V_{\text{HXC}} \;=\; \sum_j v_j \hat{n}_j \,
%\end{equation}
also including the Hartree term.
Here, $n_j$ denotes the particle density. The sum is over the lattice 
sites $\ell$ of $e\mathcal{M}$ and the states $k$
of $\mathcal{R}$. According to theorems by Kohn and Hohenberg\cite{Hohenberg_Kohn:1964}
and Kohn and Sham\cite{Kohn_Sham:1965}
a unique set of coefficients $v_j$ specifying the XC kernel exists, such 
that the KS-particle density coincides with the exact density of the
many body GS. In practice, we find $v_j$ solving a standard optimization problem, 
which yields a final relative density mismatch of less than $10^{-10}$ per site. 
%

%\paragraph{Kubo conductances}
%
Within DMRG the linear conductance can be obtained from evaluating the 
Kubo-formula \cite{Bohr_Schmitteckert_Woelfle:2006,Bohr_Schmitteckert:2007}
\begin{equation} \label{Eq:Kubo}
  g \;=\; \frac{8 \pi \e^2}{h} \bra{\Psi_0}
    \hat{J}_{n_1}\frac{\eta(\HH-E_0)}{\big[(\HH-E_0)^2+\eta^2\big]^2}
  \hat{J}_{n_2} \ket{\Psi_0},
\end{equation}
where $\ket{\Psi_0}$ is the many body ground state,
$\eta$ is the broadening parameter.
$J_n$ is the current density operator at the bond between
site $n$ and $n-1$ and $E_0$ denotes the GS energy.
Due to particle number conservation,
the $dc$-conductance is independent of $n_1,n_2$.
For the details of the procedure see Ref.~\cite{Bohr_Schmitteckert_Woelfle:2006,Bohr_Schmitteckert:2007}.  

The conductance $g$ -- as in fact any dynamical correlator
  at $T{=}0$ -- can be calculated 
  evaluating proper GS matrix elements of certain known many body
  operators, see e.~g. Eq. (\ref{Eq:Kubo}). Therefore, the general
  principles of DFT apply and functionals exists, parameterized by
  $\eta$ (or frequency $\omega$, times $t,t'$ etc.), which allow to
  calculate such correlators from the GS density $n_0$ alone. The
  functional, which would yield the exact conductance $g[n_0]$ is not
  known. An approximative expression for $g$ coinciding with the exact 
KS conductance $g\s$ 
is readily obtained employing
Eq.~(\ref{Eq:Kubo}) using the KS ground state with KS single 
particle energies $\epsilon_p$, the corresponding orbitals for evaluating
the matrix elements $J_{0p}$ and $\HH{\rightarrow}\HH\s$
\begin{equation} \label{Eq:KuboFF}
	g\s \;=\; \frac{8\pi \e^2}{h} \sum_{p,q} \frac{ J_{0p} J_{q0} 
\,\eta \left(\epsilon_p - \epsilon_q\right)}{ \left((\epsilon_p -
  \epsilon_q)^2 + \eta^2\right)^2}\,  
	f(\epsilon_q) \left( 1 - f(\epsilon_p) \right). 
\end{equation}
$f(\epsilon)$ denotes the Fermi-Dirac occupation numbers.

\begin{figure}[t]
\begin{center}
    \includegraphics[width=0.475\textwidth]{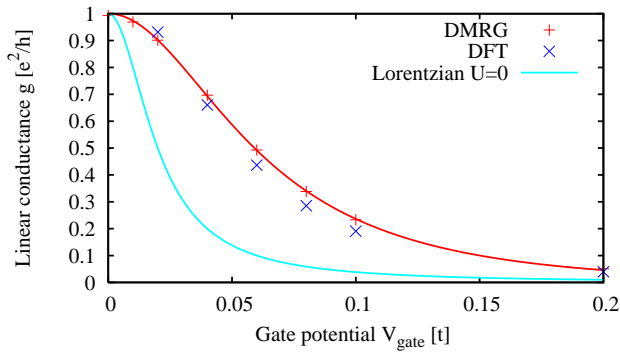}
    \caption{\ColorOnline Linear conductance over the gate voltage
 for the IRLM model with $t'=0.1$.
      Comparison of the conductance calculated  within DMRG
    for the full problem ($+$) and 
 %%%from evaluating the Kubo formula 
   for the corresponding effective DFT Hamiltonian ($\times$). 
   The line through the DMRG data is a guide to the eyes and the
   Lorentzian is the non-interacting result as reference.
   DMRG half width: $\Gamma=0.116 \pm 0.001$; DFT: $0.106 \pm 0.002$;
   noninteracting system, $U=0$: $0.04$.}
   \label{f1}
\end{center}
\end{figure}

%\paragraph{Results, single site:}
We have calculated conductances for molecules with only one site (single
interacting level) and with five sites. We begin with the single level model
with numerical parameters: $t'=0.1$. The model enjoys a particle hole
symmetry, so that the single
transport resonance is pinned to the band center, $E{=}0$. 
Fig.~\ref{f1} shows the exact, $g(V_\text{gate})$,
and the Kohn-Sham conductance $g\s$.
Comparison to the non-interacting limit ($U{=}0)$) exhibits
a strong (280\%) interaction driven 
enhancement of the resonance width, $\Gamma=0.116$, 
compared to the non-interacting case $\Gamma=4t'^2=0.04$.
Exact DFT, $g\s$, is able to reproduce
this renormalization effect with accuracy better
than 10\%, $\Gamma^\text{DFT}{=}0.106$.%

%\paragraph{Results, five sites:}
We now turn to the five site case, which affords four single levels, that are
not pinned to zero energy. Since now resonances experience
an occupation dependent and interaction driven shift (``Coulomb
blockade'') with a corresponding change of the resonance width,
this model can serve to investigate the DFT handling of such
renormalization phenomenona. 

\begin{figure}[tb]
\begin{center}
    \psfrag{Dot}{{$\cal \scriptstyle M$}}
    \includegraphics[width=0.475\textwidth]{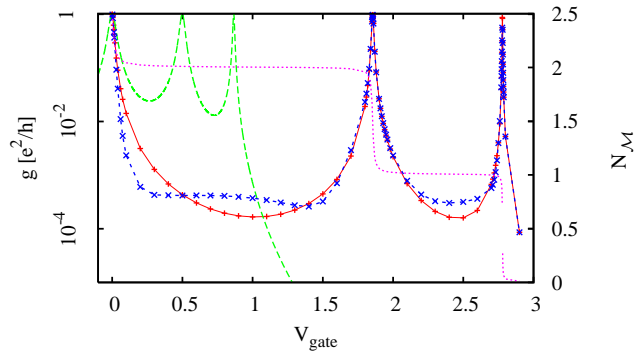}
    \caption{\ColorOnline Comparison of the exact conductance ($+$)
     and the ground state DFT approximation ($\times$, dashed line as
     a guide) for a five site system  ($t'{=}0.2$).
     For comparison the noninteracting $U{=}0$) is shown as well
     (long dashed line). Solid line indicates the particle number $N_{\cal M}(V_\text{gate})$
     the molecule. The resonance of $g$ are sitting
     $V_\text{gate}=0$, $1.854$, and $2.779$ with a resonance width of
	$\Gamma=0.026$, $0.015$, and $0.0033$. }\label{Fig:DFT5_All}
%%Gamma0: 0.0257283295495539
%%Gamma1: 0.0150493217516619  x: 1.85474058238378
%%Gamma2: 0.00325741629318578 x: 2.7786681311291
\end{center}
\end{figure}

Fig.~\ref{Fig:DFT5_All} displays the $g(V_\text{gate})$ and 
$g\s(V_\text{gate})$ conductance. 
Since particle-hole symmetry implies invariance under 
$V_\text{gate}\leftrightarrow -V_\text{gate}$, only the positive branch is shown.
%(numerical parameters: $t{=}0.5, U{=}2, t'{=}0.2$.)
The first resonance at nonzero energy signalizes the transition, where the electron
number $N_{\cal M}(V_\text{gate})$ of the molecular dot changes between 
two and one, see Fig.~\ref{Fig:DFT5_All}. 
This happens at {$V_\text{gate}\sim U{-}\Delta$}, where $\Delta$
denotes the single particle level spacing. This expectation is roughly consistent with 
the numerical value 1.8 obtained from Fig.~\ref{Fig:DFT5_All}, for details see 
\cite{Bohr_Schmitteckert_Woelfle:2006}.

Fig.~\ref{Fig:DFT5_All} clearly shows, that the DFT calculation 
perfectly well captures the position of the transport resonances.  
In addition, the broadening of the resonance peaks is described 
reasonably well. Similar to the single level case, for the center peak width 
$10\%$ deviations have to be 
accounted for. By contrast, a logarithmic plotting is required in order to
make the relative deviations visible for the broadening of the shifted peaks.  
Remarkably, near resonances the variation of the conductivity is described over more than
three orders of magnitude with deviations of a few percent or less.

% \begin{figure}[thb]
% \begin{center}
%     \subfigure[]{\label{Fig:DFT5_p1}
%     \includegraphics[width=0.475\textwidth]{DFT5a.eps}}
%     \hspace{0.02\textwidth}
% %    \subfigure[t'=0.03]{\label{G_0.03}
%     \subfigure[]{\label{Fig:DFT5_p2}
%     \includegraphics[width=0.475\textwidth]{DFT5b.eps}}
%     %
%      \subfigure[]{\label{Fig:DFT5_p3}
%     \includegraphics[width=0.475\textwidth]{DFT5c.eps}}
%     %
%    \caption{\ColorOnline On resonance comparison $\ldots$. }\label{Fig:DFT5_Peaks}
% \end{center}
% \end{figure}

%\paragraph{Discussion:}

\begin{figure}[tb]
\begin{center}
    \includegraphics[width=0.475\textwidth]{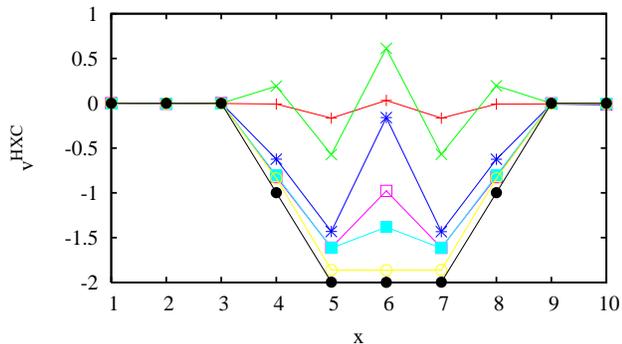}
    \caption{\ColorOnline Potential $v^\text{HXC}_\ell$ on sites $\ell$ of $e\!\mathcal{M}$ 
      corresponding to conductance data shown in previous Fig. \ref{Fig:DFT5_All} 
      at $V_\text{gate}{=}0.2,0.005,1.5,1.85,1.86,2.5,2.8$
      (labeling center site, $x{=}6$, from top to bottom).
  }\label{Fig:DFT5_KSR}
%%Gamma0: 0.0257283295495539
%%Gamma1: 0.0150493217516619  x: 1.85474058238378
%%Gamma2: 0.00325741629318578 x: 2.7786681311291
\end{center}
\end{figure}

In Fig.~\ref{Fig:DFT5_KSR} we show the evolution of the local on site potential
$v^\text{HXC}$ with increasing gate voltage. The overall behavior is
complicated, and a detailed discussion has to be relegated to 
Ref. \cite{schmitteckert07}. Here, we can only briefly 
comment on two crucial aspects. First, $v^\text{HXC}$ partially 
compensates $V_\text{gate}$ for repulsive 
voltages inbetween two resonances, keeping $N_\mathcal{M}$ integer, 
c.~f. Fig. \ref{Fig:DFT5_All}. Second, at the resonance,
$V_\text{gate}{\approx}1.855$, the center peak of Fig. 
\ref{Fig:DFT5_KSR} rapidly decays. Thus, the double well
structure in the full effective potential $v^\text{HXC}{+}V_\text{gate}$, 
that was appropriate for two repulsive particles $N_\mathcal{M}{=}2$, 
transmutes into a single well hosting the lone particle, $N_\mathcal{M}{=}1$.

The predictive power of conductance calculations with ground state
DFT may seem surprising at first sight, because $V_\text{XC}$ is
neglected albeit it is very well known that the bare KS-response 
yields incorrect excitation energies, which are shifted
to proper values in TDDFT only by including dynamical 
correlations.\cite{stefanucci07} 
A closely related fact: the bare KS-spectral
function, $A\s(\omega,V_\text{gate}=0)$  exhibits excitation peaks at 
frequencies $\omega$ of the order of $\Delta$, and not $U$.
The point, that we wish to make here is, that for correlated electron
systems the dependency of $A\s$ on its arguments $\omega$ and $V_\text{gate}$ is quite
different. The linear transport
probes $A\s$ only in the vicinity of zero frequency. The evolution
of $A\s(\omega=0,V_\text{gate})$ with gate voltage is closely tied to the 
particle number and therefore can be physically meaningful and 
give quantitative results even if dynamical corrections are ignored. 

We elaborate on this issue and give two reasons, why in fact 
the good performance of $g\s$ is {not} entirely unexpected. 
(1) We are concerned with isolated resonances $\Gamma\ll \Delta,U$,
which is a situation typical of the relatively small molecules that one deals with in the field of
Molecular Electronics. These resonances occur precisely at the 
degeneracy point, where the $N$ and $N{+}1$ particle states of the molecular dot
coincide in energy, so that the particle number is half integer
$N{+}1/2$. Since the exact DFT monitors the true particle number on the dot, the
degeneracy point of the KS-occupation and hence the KS-transport resonance
coincide with the true value. 
(2) It is much less obvious, why also the resonance width $\Gamma$ 
should always be given very accurately, and in fact there is no
reason to believe that this is the case. However, under fairly general
assumptions, one may argue that the width $\Im \Sigma$ 
of the transition in $N_{\cal M}(V_\text{gate})$ gives a very good estimate for the width 
$\Gamma$ of the transport resonance. 
%%%%%%%%%%%%%%%%%%%%%%%%%%%%%%%%%%%
Indeed, consider a molecular dot 
invariant under exchange of left and right reservoirs (symmetric coupling). 
Then the single particle lifetime proper, $\Im \Sigma^{-1}$, of molecular excitations  
also appears as a transport rate $\Gamma\approx \Im\Sigma$, since the escape rates into the 
left and right leads, $\Gamma_{L}$ and $\Gamma_{R}$,
simply coincide with $\Im \Sigma/2$. Therefore $\Gamma^{-1}$ sets the (only) 
time scale for relaxation processes and hence it should also 
describe the width of transport resonances. 
%%Actually, it is
%%for this reason that it can frequently be sufficient to focus on 
%%spectral functions even if transport properties are addressed
%%(c.f. Ref. \onlinecite{MeirWingreen92}). 
Notice, that $N_{\cal M}(V_\text{gate})$ cannot easily distinguish two situations, where 
$\Gamma_L$ and $\Gamma_R$ are vastly different with 
the sum, $\Gamma{=}\Gamma_L{+}\Gamma_{R}$,
being kept fixed. The excellent agreement found in this
work for the symmetric case may no longer pertain into the strongly asymmetric
limit, since $g\s\propto \Gamma_{L}\Gamma_{R}$.

%\paragraph{Conclusion:}
In summary, we have presented a method for performing exact DFT calculations for model
systems based on the density matrix renormalization group (DMRG). 
The approach has been used in order to calculate 
ground state Kohn-Sham conductances for the interacting resonant level model (IRLM),
which can be compared to exact results obtained with DMRG. 
We find that DFT calculations can describe positions and broadenings of 
transport resonances with a very good accuracy. In fact, 
the resonance position will be given very precisely for 
any correlated electron system, as long as it is connected 
to a single resonant free fermion level by adiabatically switching on the interaction
(Fermi liquid regime). 
A further implication suggested by our result is
that dynamical corrections should be small as long as vertex
corrections can be ignored. So the
most pressing limitations in practical DFT conductance calculations
appears to be the missing deriviative discontinuity. This poses a 
notoriously difficult problem which, however, in principle has been well understood.%
\cite{KieronsBook}

Finally, we mention that our method DFT$\oplus$DMRG is a very general
approach, and not restricted to the IRLM employed in this work.  
It will an intriguing question to be addressed in future work, to
what extend DFT can capture also those phenomena -- be it in the density
response or in transport signatures -- which live beyond the regime of 
attraction of the Fermi liquid fixed point. 

\acknowledgments Useful discussions with K.~Burke, F.~Furche,
G.~Schneider and
P.~W\"olfle are greatfully acknowledged. This work was supported by the Center of
Functional Nanostructures at Karlsruhe University.

\end{document}